\documentstyle[12pt,epsfig]{article}
\begin{document}

\begin{titlepage}
\rightline{May 2004}
%\rightline{astro-ph/04xxxx}
\vskip 3cm
\centerline{\large \bf 
Reconciling the positive
DAMA annual modulation signal
}
\vskip 0.3cm
\centerline{\large \bf 
with the negative results of the
CDMS II experiment}

\vskip 2.2cm
\centerline{R. Foot\footnote{
E-mail address: foot@physics.unimelb.edu.au}}

\vskip 0.7cm
\centerline{\it School of Physics,}
\centerline{\it University of Melbourne,}
\centerline{\it Victoria 3010 Australia}
\vskip 2cm
\noindent
We examine the recent CDMS II
results in the context of the mirror matter
interpretation of the DAMA/NaI experiment.
We find that the
favoured mirror matter interpretation of the
DAMA/NaI experiment -- a $He'/H'$ dominated halo
with a small $O'$ component is fully consistent
with the null results reported by CDMS II. 
While the CDMS II experiment is quite sensitive to a
heavy $Fe'$ component, and may yet find a positive
result, a more
decisive test of mirror matter-type dark matter
would require a lower threshold experiment using light
target elements.

\end{titlepage}

%%% start here %%%
The CDMS II experiment has
recently announced first results from their cryogenic
dark matter search in the Soudan Underground Laboratory\cite{cdms}.
With 19.4 kg-days of Ge effective exposure after cuts, 
and a threshold energy of
10 keV, they obtain strong limits on WIMP dark matter -- ruling
out a significant range of supersymmetric models. 
Furthermore, the CDMS II experiment is inconsistent with the 
impressive $6.3\sigma$ DAMA annual modulation signal\cite{dama}
if interpreted in terms of standard (spin independent) WIMPs.
It is reasonable to conclude that the standard WIMP hypothesis is
disfavoured by the experiments.

However, it has already been pointed out that a better
explanation for the impressive DAMA/NaI 
annual modulation signal
is given by mirror matter-type dark matter\cite{footdama,footdama2}.
The purpose of this note is to examine the 
latest CDMS results in the context of the mirror matter
interpretation of the DAMA annual modulation
signal.

Recall, mirror matter is predicted to exist if nature exhibits
an exact unbroken mirror symmetry\cite{flv} (for
reviews and more complete set of references, see Ref.\cite{review}). 
For each type of ordinary
particle (electron, quark, photon etc) there is a mirror partner
(mirror electron, mirror quark, mirror photon etc), 
of the same mass. The two sets of particles form 
parallel sectors each with gauge symmetry $G$
(where $G = SU(3) \otimes SU(2) \otimes U(1)$ in the 
simplest case)
so that the full gauge group is $G \otimes G$.
The unbroken mirror symmetry maps
$x \to -x$ as well as ordinary particles into mirror
particles. Exact unbroken time reversal symmetry
also exists, with standard CPT identified as the product
of exact T and exact P\cite{flv}.

Mirror matter is a rather obvious candidate 
for the non-baryonic dark matter in the Universe because:
\begin{itemize}
\item
It is well motivated from fundamental physics
since it is required to exist if parity and time reversal
symmetries are exact, unbroken symmetries of nature.
\item
It is necessarily dark and stable. Mirror baryons have
the same lifetime as ordinary baryons and couple to mirror
photons instead of ordinary photons.
\item
Mirror matter-type dark matter can provide a suitable framework for
which to understand the large scale structure of the 
Universe\cite{comelli}.
\item
Recent observations from WMAP\cite{wmap} and other experiments suggest
that the cosmic abundance of non-baryonic dark matter is
of the same order of magnitude as ordinary matter $\Omega_{dark} 
\sim \Omega_{b}$. A result which can naturally occur if
dark matter is identified with mirror matter\cite{fvwmap}.
\end{itemize}

Ordinary and mirror particles interact with each
other by gravity and via the photon-mirror
photon kinetic mixing interaction:
\begin{eqnarray}
{\cal{L}} = {\epsilon \over 2} F^{\mu \nu} F'_{\mu \nu} 
\label{km}
\end{eqnarray}
where $F^{\mu \nu}$
($F'_{\mu \nu}$) is the field strength tensor for electromagnetism
(mirror electromagnetism)
\footnote{Given the 
constraints of gauge invariance, renomalizability and mirror
symmetry it turns out\cite{flv} 
that the only allowed non-gravitational interactions
connecting the ordinary particles with the mirror particles
are via photon-mirror photon kinetic mixing 
and via a Higgs-mirror Higgs quartic
coupling, ${\cal{L}} = \lambda \phi^{\dagger} \phi \phi'^{\dagger}
\phi'$. If neutrinos have mass, then ordinary - mirror
neutrino oscillations may also occur\cite{flv2}.}. 
Photon-mirror photon mixing causes 
mirror charged particles to couple to 
ordinary photons with a small effective
electric charge, $\epsilon e$\cite{flv,hol,sasha}, leading
to many interesting implications for astrophysics, particle
physics and related fields\cite{app}.

One such interesting implication [of Eq.\ref{km}] 
is that the 
DAMA/NaI experiment is sensitive to mirror matter-type dark
matter\cite{footdama}.
Halo mirror atoms can elastically scatter off ordinary atoms as
a consequence of
the photon-mirror photon kinetic mixing interaction, Eq.(\ref{km}).
The DAMA experiment is not particularly sensitive to very
light dark matter particles such as mirror hydrogen and mirror
helium. Impacts of these elements (typically) do not
transfer enough energy to give a signal above the detection
threshold\cite{footdama}. The next most abundant element is expected to
be mirror oxygen (and nearby elements). 
A small mirror iron component is also possible.
Interpreting the DAMA annual modulation signal in terms
of $O', Fe'$, we found that\cite{footdama}:
\begin{eqnarray}
|\epsilon | \sqrt{ {\xi_{O'} \over 0.10} +
{\xi_{Fe'} \over 0.026}} 
\simeq 4.8^{+1.0}_{-1.3} \times 10^{-9}
\label{dama55}
\end{eqnarray}
where the errors denote a 3 sigma allowed range and 
$\xi_{A'} \equiv \rho_{A'}/(0.3 \ {\rm GeV/cm^3})$ 
is the $A'$ proportion (by mass) of the halo dark matter.

%In this theory the halo consists of compact
%objects such as mirror stars and planets, comprising
%about 25\% of the halo (by mass), with the remaining 75\% being
%composed of mirror atoms/nuclei.
%The chemical composition of the atoms/nuclei can
%be approximated by four elements: $H', He', O', Fe'$.

In Ref.\cite{footdama} we found that a DAMA/NaI
annual modulation signal dominated by the $Fe'$ component,
is experimentally disfavoured for three independent reasons:
a) it predicts a differential energy spectrum rate
larger than the measured DAMA/NaI rate 
b) potentially leads to a significant diurnal effect c) should
have been observed in the CDMS I experiment.
Thus it is probable that the
mirror oxygen component dominates the 
DAMA annual modulation signal, which from Eq.(\ref{dama55})
means that $\xi_{O'} \stackrel{>}{\sim} 4\xi_{Fe'}$.
In this case there are no significant problems with
existing experiments.

If the DAMA signal is dominated by $O'$, then things
depend on only one parameter, $\epsilon \sqrt{\xi_{O'}}$.
This parameter is fixed from the annual modulation
signal, Eq.(\ref{dama55}), which means that the
event rate (due to $O'$ interactions) can
be predicted for other experiments.
The prediction does depend on the assumed halo
distribution.
In our analysis we assume that the halo has the 
standard isothermal
Maxwellian distribution 
\begin{eqnarray}
f_{A'}(v)/k = (\pi v_0^2)^{-3/2} \ exp[-v^2/v_0^2] .
\end{eqnarray}
It is important to realize that the $v_0$ value
for a particular halo component element, $A'$,
depends on the chemical composition of the halo.
In general\cite{footdama2},
\begin{eqnarray}
{v_0^2 (A') \over v_{rot}^2} = {\mu M_p \over M_{A'}}
\label{z3}
\end{eqnarray}
where $\mu M_p$ is the mean mass of the particles
comprising the mirror (gas) component of the halo ($M_p$ is the
proton mass) and $v_{rot} \simeq 220$ km/s is the local rotational
velocity.
The most abundant mirror elements
should be $H', He'$, generated in
the early Universe from mirror big bang nucleosynthesis
(heavier mirror elements should be generated in mirror
stars). It is useful, therefore, to consider 
two limiting cases: first that the halo
is dominated by $He'$ and the second is that
the halo is dominated by $H'$. The $v_0$ values can
be easily obtained from Eq.(\ref{z3}), taking into
account that the light halo mirror atoms should be fully ionized:
\begin{eqnarray}
v_0 (A') &=& v_0 (He') \sqrt{{M_{He'}\over M_{A'}}} \approx {220 \over
\sqrt{3}}
\sqrt{{M_{He'}\over M_{A'}}} \ km/s \ \  \ \rm{for \ He' \ dominated \ halo}\nonumber \\ 
v_0 (A') &=& v_0 (H') \sqrt{{M_{H'}\over M_{A'}}} \approx {220
\over \sqrt{2}}
\sqrt{{M_{H'}\over M_{A'}}} \ km/s \ \ \ \rm{for \ H' \ dominated \ halo}.
\end{eqnarray}
Mirror BBN\cite{comelli}
suggests that $He'$ dominates over $H'$, but
we will consider both limiting cases in this paper. 

It is a straightforward exercise to work out the
predicted event rate for the CDMS II/Ge and CDMS II/Si
experiments. 
[See Ref.\cite{footdama} for details of the cross section, form
factors and rate equations used].
In {\bf figure 1} we give the predicted event
rate for the CDMS II/Ge experiment
taking into account the published detection efficiency
(figure 3 of Ref.\cite{cdms}).
{\bf Figure 2} is the corresponding figure for the CDMS II/Si experiment
(assuming the same detection efficiency).
As the figures show, the event rate is predicted to be very low.
For the CDMS II/Ge experiment
the predicted event rate is just 1 event per $2.6\times 10^6$ kg-days
for $He'$ dominant halo and 1 event per $5\times 10^{12}$ kg-days if
$H'$ dominates the halo.
Given that CDMS II has only 52.6 kg-day raw exposure in Ge, this implies
a predicted number of events of just $2\times 10^{-5}$ (assuming $He'$ dominant halo)
and even less if $H'$ dominates the mass of the halo.
Clearly this prediction is nicely consistent with the null
result of CDMS II/Ge. 
In the case of the 
CDMS II/Si experiment, the predicted event rate (due to $O'$ interactions) 
is 1 event per 710 kg-days 
(for $He'$ dominated halo) and 
1 event per 200,000 kg-days (for $H'$ dominated halo).
CDMS II/Si currently has about 20 kg-days of raw exposure in Si,
so CDMS II/Si is also not sensitive to $O'$ dark matter.
Overall, the CDMS II experiment
is not sensitive to $O'$ dark matter --
certainly many orders of magnitude less sensitive than the DAMA/NaI
experiment.
\vskip 0.6cm
\centerline{\epsfig{file=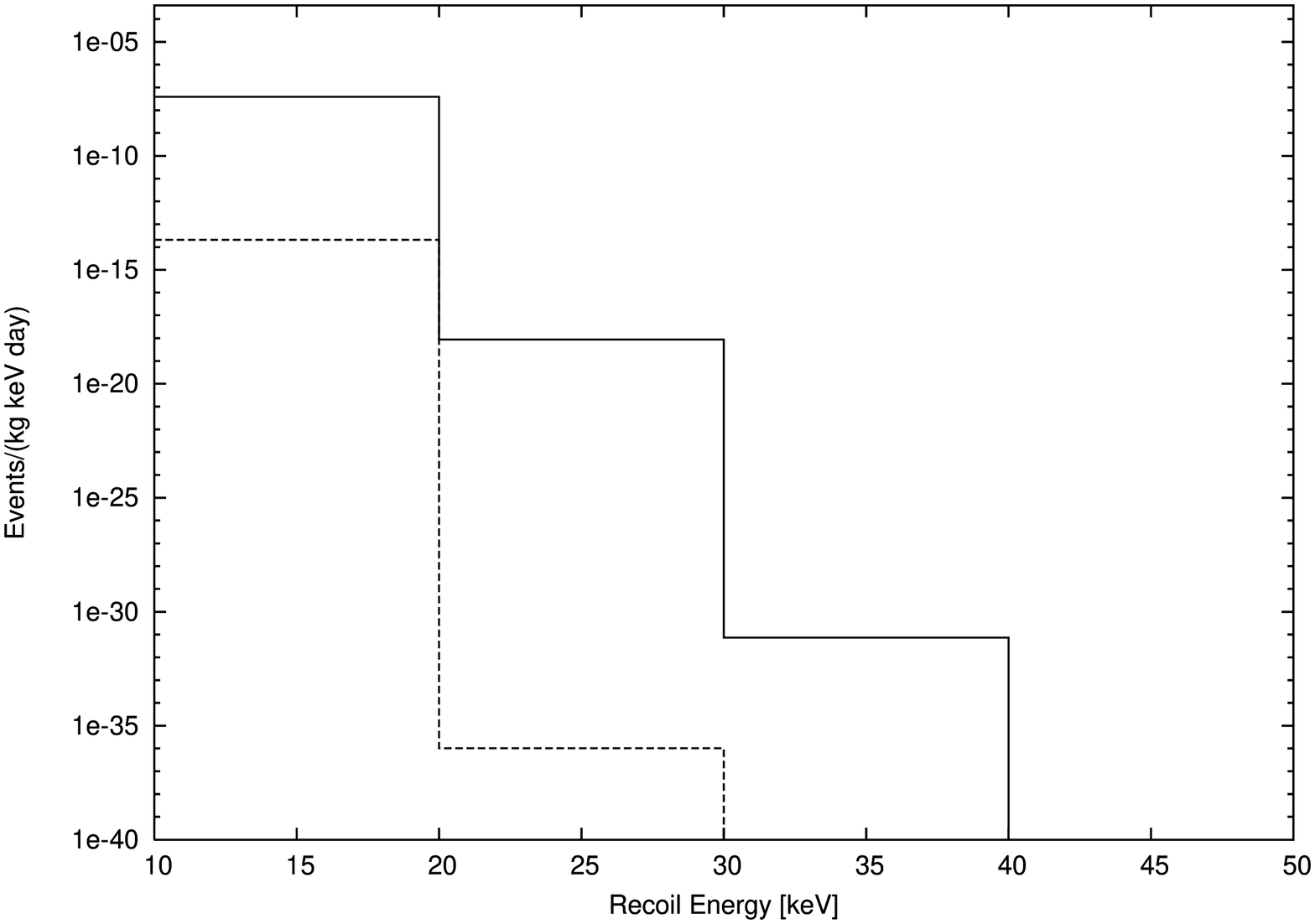,angle=270,width=12.6cm}}
\vskip 0.4cm
{\small \noindent Figure 1:
Predicted differential event rate, $dR/dE_R$, (binned into 10 keV
bins) due to $O'$ dark matter
with $\epsilon \sqrt{\xi_{O'}/0.10} = 4.8\times 10^{-9}$ 
(DAMA/NaI annual modulation best fit)
for the CDMS II/Ge experiment.
The solid line corresponds to 
a standard halo
model with $He'$ dominated halo 
while the dashed line assumes
a $H'$ dominated halo.
}
\vskip 0.5cm
\centerline{\epsfig{file=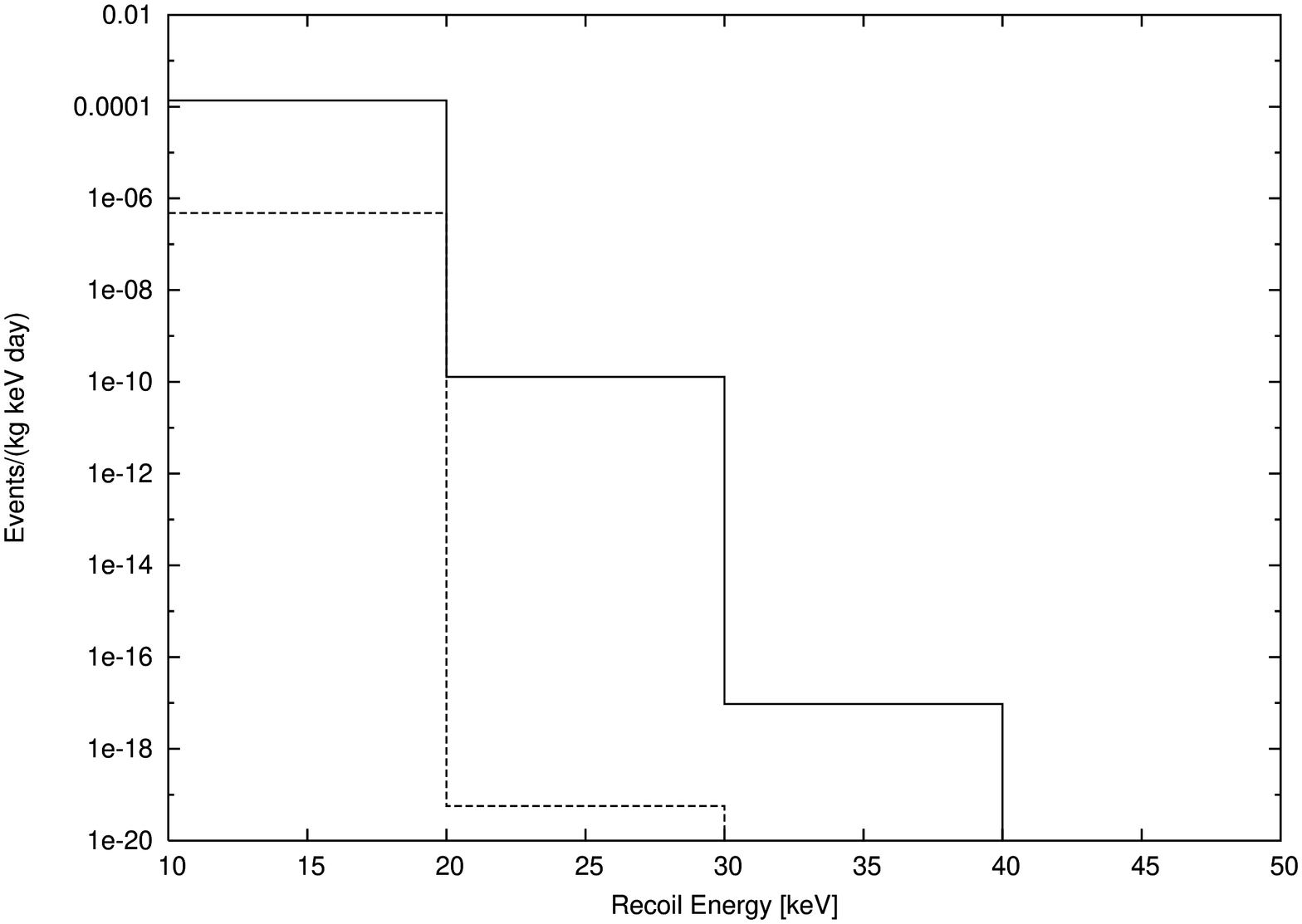,angle=270,width=12.6cm}}
\vskip 0.4cm
{\small \noindent Figure 2:
Same as Figure 1, except for the CDMS II/Si experiment.}
\vskip 1.0cm

In the case of standard spin independent WIMPs, the
CDMS II experiment is more sensitive 
than the DAMA/NaI experiment. However, as we have discussed
above, this is clearly not the case for $O'$-type
dark matter (with dominant $He'/H'$ component).
The diverse behaviour of the two types of dark matter
candidate has to do with their basic differences:
\begin{itemize}
\item
The differential cross section for
mirror matter-type dark matter is inversely
proportional to the square of the recoil energy, while
that for WIMPs is energy independent (excepting the
energy dependence of the form factors). 
\item 
For $O'$-type dark matter in an $He'/H'$ dominated halo,
$v_0 (O') \ll 220$ km/s, while the characteristic 
velocity of WIMPs are assumed to be approximately 220 km/s.
\item 
The mass of $O'$ is only 15 GeV, while WIMPs
are typically constrained from W,Z decays to be heavier than
$30-45$ GeV (depending on the model).
\end{itemize}
These three differences mean that experiments with
low threshold energy and light target elements
are much more sensitive (to $O'$-type dark matter) than experiments
with higher threshold energy and/or heavy target
elements. In the case of DAMA/NaI, the event
rate for mirror matter-type dark matter
is dominated by interactions 
with the light $Na$ component. The actual threshold energy
of 6.7 keV (for $Na$), implies a threshold impact velocity of
290 km/s. In the case of CDMS II/Ge, the threshold energy
of 10 keV and heavy Ge target gives a threshold impact
velocity of 450 km/s (see Ref.\cite{fd3} for a table
of threshold velocities for the various experiments). Given the
low value of $v_0 (O')$ [$v_0 (O') = {110 \over \sqrt{3}}$ km/s 
(${55 \over \sqrt{2}}$ km/s) for $He'$ 
($H'$) dominated halo] the number of $O'$ atoms with impact velocity
above threshold is clearly much lower for CDMS II/Ge compared with
DAMA/NaI.
Note that the Edelweiss I/Ge and Zeplin I/Xe experiments
are even less sensitive than CDMS II/Ge because the threshold
impact velocity of those experiments is even higher\cite{fd3}.

Although the CDMS II experiment is relatively insensitive to
impacts of $O'$, it is much more sensitive to
impacts of heavier mirror atoms.
If we assumed that the DAMA/NaI experiment
were due to $Fe'$ interactions in a $He'/H'$ dominated halo,
then we would have predicted around 30 events 
(roughly independently on whether $H'$ or $He'$ dominates the mass
of the halo) for
the CDMS II 52.6 kg-day Ge raw exposure.
This is clearly excluded by the data which failed
to find any events. Evidently $Fe'$ can only
be a small component of the halo, with energy density
much less than $O'$. Assuming the standard halo
model, we find that
$\xi_{Fe'} \stackrel{<}{\sim} \xi_{O'}/40$ at
90\% C.L. (independently of
whether $He'$ or $H'$ dominates the mass of the halo).
Clearly, future CDMS data may well find a positive
signal -- corresponding to an $Fe'$ component --
which should be there at some level.

In conclusion, we have examined the recent CDMS II
results in the context of the mirror matter
interpretation of the DAMA/NaI experiment.
The favoured mirror matter interpretation of the
DAMA/NaI experiment -- a $He'/H'$ dominated halo
with a small $O'$ component is fully consistent
with the null results reported by CDMS II. 
While the CDMS II experiment is quite sensitive to a
heavy $Fe'$ component, and may yet find a positive
result, a more
decisive test of mirror matter-type dark matter
would require a lower threshold experiment using light
target elements (such as DAMA/LIBRA and CRESST II).

\vskip 1cm
\noindent
{\bf Acknowledgements}
\vskip 0.5cm
\noindent
This work was supported by the Australian Research Council. 

\end{document}